# Xenon plasma-focused ion beam milling as a method to deterministically fabricate bright and high-purity single-photon sources operating at C-band

MACIEJ JAWORSKI,[1,2] PAWEŁ MROWIŃSKI,[1] MAREK BURAKOWSKI,[1] PAWEŁ HOLEWA,[1,3,4] LAURA ZEIDLER,[1] MARCIN SYPEREK,[1] ELIZAVETA SEMENOVA,[3,4] GRZEGORZ SĘK[1]

[1] *Department of Experimental Physics, Faculty of Fundamental Problems of Technology, Wrocław University of Science and Technology, Wybrzeże Wyspiańskiego 27, 50-370 Wrocław, Poland*
[2] *Nanores, Bierutowska 57-59, 51-317 Wrocław, Poland*
[3] *DTU Electro, Technical University of Denmark, Kongens Lyngby 2800, Denmark*
[4] *NanoPhoton-Center for Nanophotonics, Technical University of Denmark, 2800 Kongens Lyngby, Denmark*

**Abstract:** Electron beam lithography is a standard method for fabricating photonic nanostructures around semiconductor quantum dots (QDs), which are crucial for efficient single and indistinguishable photon sources in quantum information processing. However, this technique lacks direct 3D control over the nanostructure shape, complicating the design and enlarging the 2D footprint to suppress in-plane photon leakage while directing photons into the collecting lens aperture. Here, we present an alternative approach to employ Xenon plasma-focused ion beam (Xe-PFIB) technology as a reliable method for the 3D shaping of photonic nanostructures containing low-density self-assembled InAs/InP quantum dots emitting in the C-band range of the 3rd telecommunication window. We explore both deterministic and non-deterministic fabrication approaches, resulting in mesas naturally shaped as truncated cones. As a demonstration, we fabricate mesas using a heterogeneously integrated structure with a QDs membrane atop an aluminum mirror and silicon substrate. Finite-difference time-domain (FDTD) simulations show that the angled sidewalls significantly increase photon extraction efficiency, achieving $\eta = 0.89$ for NA = 0.65. We demonstrate experimentally a high purity of pulsed single-photon emission (~99%) and a high extraction efficiency of $\eta = 0.24$, with the latter surpassing the highest reported values obtained using electron beam lithography in the C-band.

## 1. Introduction

In recent years, scientists worldwide have focused their efforts on studying single photon sources and their properties for emerging quantum technology applications. In this context, several competing approaches exist to generate non-classic light from solid-state systems, such as the growth of low-dimensional semiconductors [1] and the fabrication of color centers in diamonds [2] or ion traps [3]. The subject of our research are semiconductor quantum dots (QDs), which are especially promising as they demonstrate several advantageous properties when compared to other solutions, including also the laser-based single photon sources. The most important are the ability to control the emission wavelength to match the telecommunication windows [4][5][6][7], triggered and on-demand emission of single photons [8], their high generation rates [9], and possible photonic chip integration with the use of QD localization [10] and deterministic processing [11][12][13]. The on-demand generation of single photons with high emission purity and brightness is desirable for potential applications in quantum communication and quantum computing schemes [14]. They promise the realization of scalable quantum photonic integrated circuits [15] and have been demonstrated in quantum key distribution for secured communication networks [16], where low multiphoton contribution and the coherence in the emitted photon-number states provide additional security benefits [17][18].

Efficient on-demand generation of single photons requires the fabrication of photonic microstructures containing deterministically integrated QD. A proper design shall offer directionality of emission, improving the photon extraction efficiency [19], or simultaneously, as in the case of resonant cavities, acceleration of spontaneous emission rate due to the Purcell effect [20]. In addition, the ideal spectral range of single-photon emitters for applications in optical fiber networks would be the telecom C band due to the minimum of the optical signal attenuation. The main material system candidates are InAs QDs grown on InP, which have been proven to be easily tunable to the wavelength of 1.55 μm [21][6][22][23]. The commonly used approach for fabricating photonic structures employs electron beam lithography (EBL) combined with etching, which, however, has some drawbacks. For instance, the EBL processing requires using electron resists layers deposition very often still combined with conventional masks, process, which complicates the processing. Although even more advanced in-situ EBL in combination with cathodoluminescence technique allows shaping microstructures like monolithically integrated microlenses [24], compared to that the FIB milling is a one-step process which is a significantly simpler technology [25]. Therefore, an interesting alternative is based on using a double-beam Scanning Electron Microscope/Focused Ion Beam (SEM/FIB) system to process photonic microstructures, which has already been proven successful for GaAs-based quantum dot structures [26]. Here, we extended the capabilities of this technology to more application-relevant systems with InAs/InP QDs as the emitters, integrated additionally with the silicon platform, towards efficient, telecom single photon sources obtained deterministically in a simple and cost-effective technology.

The SEM/FIB system contains an electron beam column for high-resolution imaging and an ion beam column for three-dimensional milling. Typically, this configuration is used in the sample preparation process for Transmission Electron Microscopy (TEM) or failure analysis [27][28], as this is a maskless and direct method, which makes it optimal for prototyping in a fast and cost-effective manner. Additionally, this method can be used to fabricate photonic microstructures around quantum emitters, although the negative effect of incident ions causing defects in the crystal structure along incident and lateral dimensions must still be considered [29]. To counteract the negative effects of the ion beam on the crystal lattice, the demonstrations have concerned so far mostly QDs placed in the micro-cavity defined by the distributed Bragg reflectors (DBR), which at the same time protected the dots against the ion bombardment by a multilayer material with a total thickness of above 1.5 μm [30][31]. On the other hand, using DBR-based cavity structures is makes the growth complex and expensive. Although such configuration is able to offer high extraction efficiency, its disadvantage is spectrally narrow photonic mode, which requires very precise spectral matching with the QD emission lines. On the other end, there are photonic confinement structures in a form of mesas or microlenses, which use much thinner capping and offer a good compromise between the maximal extraction efficiency and its spectral broadening, which is at least an order of magnitude larger than for microcavities. Therefore, our focus revolved around structures with thinner cap layers, which simplifies the growth technology and simultaneously allows obtaining spectrally broad extraction efficiency, especially advantageous when dealing with strongly inhomogeneous QD ensembles [29][26].

Mesas were processed by the Xe-PFIB. The Xe ions are indifferent with the lack of electron affinity to III/V elements [32][33]. On the other hand, the Xe-PFIB technique increases surface amorphization due to the high redeposition rate of the material, however, the resulting roughness is composed of artifacts of below 40 nm size, which does not have a detrimental effect on the final shape and size of the few microns large structures [26]. Additionally, the beam spot size for the xenon source is greater than for gallium and depends strictly on the energy and current of the beam. For example, for the 1 nA beam current, the spot size ($\varphi$) is expected to be at most 150 nm large, and together with fixed overlap ($\sigma$) of the successive spots of 50% [34], gives that the milled area is uniformly exposed to the beam ions[35].

This study demonstrates both non-deterministic and deterministic technology leading to the fabrication of micrometer-size photonic nanostructures in the form of mesas containing self-assembled QDs used as non-classical photon states generators for quantum technology applications. The mesas are defined in a one-step process by the Xenon Plasma Focused Ion Beam (Xe-PFIB), allowing for fast (2.5 min), nanometer-precision milling of mesa structures around the optically located InAs/InP QDs, emitting in the telecom C-band at 1560 nm. We show that the precise placing of a QD in a mesa combined with the presence of the metallic mirror beneath the mesa leads to bolstering photon source brightness by increasing emission directionality due to the photonic confinement giving a spectrally broad photon extraction efficiency function with the experimentally obtained $\eta \sim 40\%(11\%)$ at NA = 0.65(0.4). The fabricated photon source revealed high photon generation purity with $g^2(0) = 8 \times 10^{-3}$ and triggered operation under driving by a sequence of optical pulses with 76 MHz frequency.

## 2. Fabrication – QD structure and Xe-PFIB processing

The QD sample was grown by metalorganic vapor phase epitaxy (MOVPE). The self-assembled InAs/InP QDs were formed in the Stranski-Krastanow growth mode and embedded in a 580 nm-thick InP [36]. The wafer is coated by plasma-enhanced chemical vapor deposition with a 603 nm thick $SiO_2$ layer and subsequentially with 130 nm of Al from the electron-beam evaporator. The QDs wafer is then integrated using benzocyclobutene (BCB) bonding with a Si wafer. The InP substrate and buffer layer are removed in the HCl bath, whereas the InGaAs sacrificial layer is removed in (10%)$H_2SO_4$:$H_2O_2$:$H_2O$=1:8:80 solution. The bonding procedure is performed at 250°C in a vacuum under the applied mechanical force of approximately 2 kN [36]. Aiming at telecom single photon emission, we used a sample with a low density of QDs ($\sim 2.8 \times 10^9$ cm$^{-2}$) emitting in a wavelength range from 1.5 to 1.6 μm. Their morphological properties are described in more detail in Ref. [37].

The scheme of the sample structure milling is schematically depicted in Fig. 1a). The ion beam energy was set to the lowest available value of 10 keV, which minimizes the structure degradation upon ion beam exposure. Thus, the radiative recombination from quantum dots is not diminished, allowing a one-step maskless approach using no protective layers, like carbon layers, as was required in the case of GaAs-based samples [26]. Based on our previous experience [26][38], we optimized the process in terms of the fabrication time, resolution, and structural and optical quality of the milled photonic structures. We found the optimal beam current at 1 nA and dwell time of 1 μs.[26] This combination of parameters enabled to fabricate mesas within approx. 2.5 min. In Fig. 1b), we demonstrate an SEM image of the milled mesa with angled sidewalls with a bottom diameter of about D = 3.5 μm and a height of $h$ = 580 nm. We used an inner-to-outer milling mode in a basic ring pattern with fixed inner and outer diameters (the outer field diameter is about 16 μm). Furthermore, we successfully created an array of 30 mesas, which differed only in the bottom diameter (*D)* of approximately 2.0 μm, 3.5 μm, and 5.0 μm, while the top diameter (*d*) is typically lower by approx. 0.6 μm. This shape can be related to the profile of the ion beam and can be controlled by the ion beam energy and beam current to some extent.

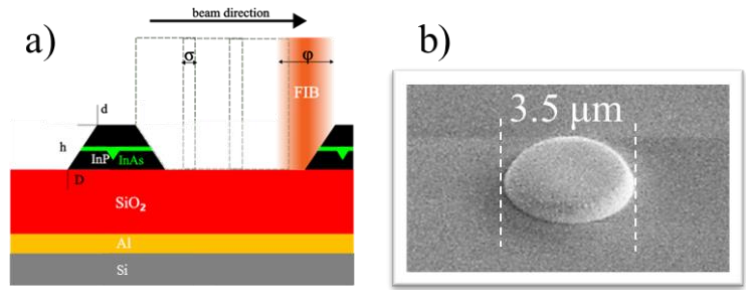

Fig. 1. a) Schematically presented layout of the sample processed by the focused ion beam. The milling is characterized by a beam diameter (φ~150 nm) and overlap (σ=50%) of the beam during milling. In this fabrication scheme, the mesa has a truncated cone shape with $d$ as the top diameter, $h$ as height, and $D$ as the bottom diameter. b) (a) Xe-PFIB-milled mesa inspected by the scanning electron microscope in the double-beam system.

In a non-deterministic approach, the microstructures are milled randomly in an array, and depending on the areal density, there is a limited probability of selecting an optimized photonic structure with a well-centered single QD. Therefore, the presented Xe-PFIB methodology has also been used deterministically, where the dot is preselected first by a photoluminescence mapping (see description in Section 4). Thus, the QD position is known, and the photonic microstructure could be fabricated at its exact position.

### 3. FDTD simulation results

The angled sidewalls in the truncated cone shape of the mesa microstructures realized by the Xe-PFIB are favorable for increased extraction efficiency due to the refraction of light towards the collecting lens [39][40][41]. We performed FDTD simulations for this specific mesa geometry to provide more insight into the case of InP-based microstructures on thick $SiO_2$ and Al mirror, focusing mainly on the light collection in the out-of-plane direction as well as on the potential Purcell enhancement provided by photonic confinement due to the increased photonic density of states.

The simulations are performed for a point-like dipole emitter situated in the center of an InP mesa, resolving the spectral range from 1.4 to 1.7 μm. The 3D model of the system simulated by FDTD is schematically shown and described in Suppl Inf. We begin with the optimization scan for the mesa bottom diameter, focusing on the extraction efficiency within 40 deg. angle cone of the detection system (40 deg. corresponds to NA = 0.65 of a microscope objective commonly used in μPL experiments, also in the measurements here – see the following subsection) for the dipole emission around 1.55 μm wavelength. The results are shown in Fig. 2a). Due to the bottom metallic mirror, we can see that for the whole range of probed sizes, the extraction efficiency ($\eta$) is expected to be not less than $\eta = 0.15$, typically much higher, reaching even $\eta = 0.89$ for a 4 μm-in-diameter mesa [42]. The presence of the pronounced maximum originates from the constant inclination of the wall as the mesa diameter is changed. The weak-cavity design in our mesas shows relatively moderate Purcell enhancement, usually not exceeding 2, i.e., significantly lower than for the high-Q cavity systems, both in-plane [43] or vertical, with top and bottom mirrors [44].

The high-Q cavities, with their ability to provide $\eta$ of more than 0.8, offer a practical solution for achieving high coupling efficiency to an external single-mode fiber via the cavity resonant mode. Despite the mesa geometry not yielding high-Q factors, we can still achieve increased extraction efficiency. This is possible due to photonic-confinement-driven emission directionality, especially when combined with high numerical aperture collection optics (microscope objective in the free space or large NA fiber directly bonded to the mesa) [7]. This approach is not limited to the narrow spectral range [41], which is one of the main advantages of such a solution. The full spectrum of results for NA = 0.65 and both $D$ = 4.0 μm and $D$ = 3.5 μm mesa size is presented in Fig. 2b). In the first case, a broadband improvement of photon extraction is expected ($\eta$ > 0.6 for more than 80 nm spectral range) with the peak maximum of $\eta$ = 0.91 at 1.543 μm. The optical field distribution in the cross-section and the far-field pattern, presented in Fig. 2c) implies improvement of extraction efficiency due to angled sidewalls, which refract the in-plane dipole emission towards the out-of-plane direction manifested by the additional side-peaks in the far-field. It is worth noting that by using a microscope objective with NA = 0.4, the "side peaks" collection is limited, resulting in 2.4 times reduced extraction at 1.55 μm (see Fig 2d). Considering the mesa of 3.5 μm in diameter corresponding to the fabricated one, the calculated value for NA = 0.65 is $\eta$ = 0.42 at 1.55 μm wavelength (for NA = 0.4, the $\eta$ is 3.8 times lower), and its spectral dependence shows a maximum ($\eta$ > 0.6) at slightly shorter wavelengths around 1.48 μm (see Fig 2b).

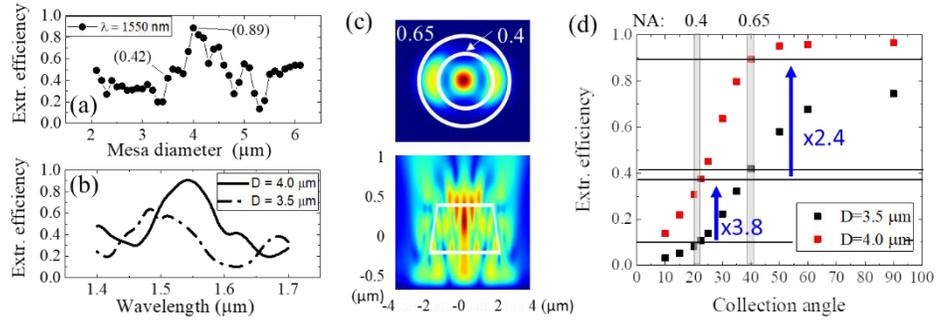

Fig. 2. (a) Calculated extraction efficiency ($\eta$, NA=0.65) for photonic microstructures at 1.55 μm wavelength varying the bottom mesa diameter, and (b) wavelength dependence of the specific mesa with 4 μm in diameter. (c) Field distributions |E| for 4 μm size mesa at 1.55 μm wavelength showing far-field distribution (upper panel) along the normal direction and cross-sectional near-field distribution in the mesa (lower panel).

## 4. Spectroscopic results

To evaluate the impact of the mesa-milling process and its influence on the QD emission properties and photon extraction efficiency, we performed high spatial resolutions photoluminescence measurements (μPL) at cryogenic temperature (7 K). For the experiment, samples were mounted on a copper cold finger in a liquid-helium-flow cryostat and excited by a continuous-wave (CW) semiconductor laser diode at 660 nm. A 0.65 numerical aperture infinity-corrected microscope objective with ×50 magnification provided the excitation and emission collection. The collected photons were spectrally analyzed in the 0.5-m focal-length monochromator using a 600-grooves/mm grating and a multichannel liquid-nitrogen-cooled InGaAs linear detector and with the tunable fiber optic filtering system. The μPL setup with monochromator gives a spectral resolution of ~25 μeV and a spatial resolution of ~2 μm, allowing for spectrally isolating single QDs and exciting a single mesa structure, while the detection based on tunable fiber optic filtering is limited by a bandwidth (full width at half maximum) of 15 μeV (0.3nm). The tunable fiber optic filtering is used for all-fiber-based Hanbury-Brown and Twiss (HBT) interferometer to probe single photon emission purity from a QD via measuring the second-order correlation function. Matching the chosen spectral line with the spectral filter, the photons enter a time-correlated single photon counting system using two NbN superconducting nanowire single-photon detectors (SNSPD). The system provides ~40 ps temporal resolution and ~60% photon detection efficiency at 1550 nm wavelength. The histograms are recorded with the time bin of 100 ps.

Near-infrared (NIR) photoluminescence (PL) imaging for QDs positioning before the FIB milling is realized in a wide-field microscope configuration with a 2D thermo-electrically cooled InGaAs camera. The 660 nm CW laser uniformly illuminates the sample surface of more than (50×50) μm² to excite the QD emission. For that, the laser beam is shaped by the beam expander and focused on the backside of the microscope objective. The PL image is collected by the same objective and transmitted through a spectral filtering system: a 1.2 μm or 1.5 μm long-pass filter followed by a bandpass filter centered at 1.55 μm with the 15 nm or 6 nm full width at half maximum. The QD localization is determined in the image post-processing, similar to the method described in more detail in Ref. [13], elaborated with an additional denoising via cross-correlation of the PL map with the 2D Gaussian peak. The setup magnifies the illuminated area 200 times, allowing the recording of the field of ~250 μm². The spatial uncertainty of the emitting point peak position is ~100 nm.

To ascertain the photon extraction efficiency ($\eta_{QD}$) of the investigated QD, we adopt the approach using tunable laser focused on the mirror mounted in the sample space [41]. The laser in the setup was tuned to the 1560 nm (within the emission range of the QD under investigation). The laser's intensity was decreased using neutral density filters (to avoid oversaturation of the detector) to attain a count rate in the megahertz (MHz) range on the superconducting nanowire single-photon detector (SNSPD). This count rate is adjusted considering factors such as the mirror's reflectivity (97%), filter attenuation, transmission through the cryostat window (90%), and the microscope objective (50%). The setup efficiency measured for a reflected laser is given for a well-focused spot which is almost ideally coupled to a single mode fiber via aspheric lens of NA=0.15. Based on the laser power incident on the silver mirror, the estimated setup efficiency is found to be $\eta_{Setup}$ = (0.36 ± 0.05) %, where the uncertainty, represented by the standard deviation σ(η), primarily arises from minor variations in the efficiency of fiber in-coupling across different wavelength. However, in the case of the mesa emission we should consider less than ideal coupling to the single mode fiber, due to mismatch of acceptance angle, i.e. etendue of the optical system related to the source(detector) diameter and the numerical appertures of collecting(focusing) lenses. In this sense, the source diameter can approximated by the mesa diameter d=3.5 μm (see broad field distribution in Fig.S1c) and the detector diameter is a fiber core diameter of d=8.2 μm, while the collection(detection) is realized via NA=0.65(0.15). The ratio of etendue ($\epsilon=S\pi NA^2$, $S=\pi d^2/4$) for these optical systems is given by $\epsilon\_det/\epsilon\_col=0.29$. Including this additional limitation for transferring photons from the mesa to SNSPD via optical fiber we end up with $\eta_{Setup,mesa} = \eta_{Setup}\epsilon\_det/\epsilon\_col = 0.1\%$. In other words, the 0.36% of setup efficiency translates to hitting the SNSPD only with the central peak observed in the far-field within the NA=0.4, while to evaluate extraction efficiency to the first lens including side peaks for NA>0.4, the setup efficiency is 0.1%.

To experimentally estimate photon extraction efficiency in the spectral range near 1.55 μm from a mesa, it is necessary to identify emission lines from a single QD in that range. The μPL results are presented in Figure 3, tackling the selected non-deterministic mesa structure with a diameter of 3.5 μm. The μPL spectra in Figure 3a), recorded using different excitation powers, give an insight into the fundamental excitonic states of a single QD emitting near the 1.55 μm wavelength within this mesa. At low CW excitation power (5 μW), the mesa response consists of two main, well-isolated, high-intensity lines. The linear polarization-resolved experiment (Figure 3b)) revealed that the line at ~1557 nm exhibits significant polarization splitting of ~74 μeV, suggesting the presence of the fine structure, typical for neutral exciton (X) in a QD with some potential confinement anisotropy [46]. With increasing the excitation power, a new μPL line appears at ~1563 nm, i.e., at lower energy - this line has the same linear polarization splitting of ~74 μeV. Moreover, the splitting phase is orthogonal to that observed for the line at ~1557 nm (Figure 3a)), which allows us to interpret the longer wavelength one as related to emission from a biexciton state (XX) of the same QD as typical for the XX-X cascaded recombination process. Hence, the XX binding energy could be estimated to be 3.3 meV, which might be considered typical for QDs of InAs on an InP system [36][45][46].

In the low-excitation μPL spectrum, the second intensive single line can be spotted at ~1566 nm, which does not exhibit any splitting in the polarization-resolved experiment (see Fig. 3b), suggesting the spin singlet nature of the recombining state linked to the charged exciton (CX) recombination. The energy separation from the neutral exciton line is approx. 4.7 meV, therefore comparable to the negatively charged exciton binding energy in an InAs/InP QD [46][47].

The assignment of the spectral lines' origin is additionally supported by the analysis of their intensity ($I$) evolution with the increasing excitation power ($P_{exc}$) presented in Figure 3c). We observe a nearly linear emission intensity increase for the X line ($I \propto P_{exc}^1$) and almost quadratic for the XX one ($I \propto P_{exc}^2$). Such intensity dependence fits a few-level kinetic rate equation model for the occupation of the X and XX states in a single QD, providing the exciton/biexciton lifetime ratio of ~2 [48][49]. At the same time, the CX line intensity evolution gives $I \propto P_{exc}^{1.2}$, with an exponent value slightly larger than one, also typical for charged excitons. These experiments provide solid arguments for identifying X, XX, and CX emissions from the same QD enclosed in the FIB-milled mesa. One can note that for elevated $P_{exc}$, more spectral features appear in the μPL spectra, especially in the 1562 -1570 nm range, contributing to a background emission for the selected XX and CX lines. The background appearance can be attributed to the recombination of higher exciton complexes in the investigated QD or can come from other QDs enclosed within the same mesa structure. However, to estimate photon extraction efficiency, the excitation power can be kept low enough to effectively eliminate the background photons from the photon number statistics.

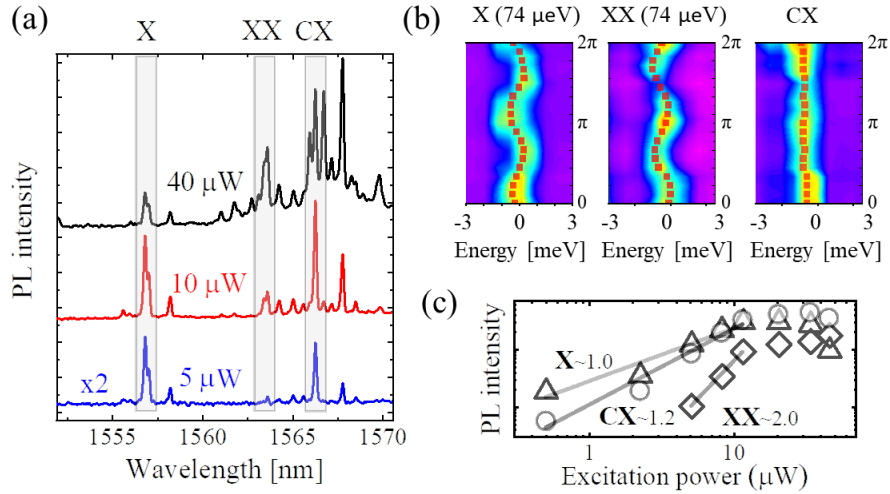

Figure 3. (a) High-spatially-resolved photoluminescence (μPL) spectra registered at different excitation power ($P_{exc}$) for the selected mesa in the spectral window near 1550 nm. (b) Spectral evolution of the X, XX, and CX lines, labeled as in (a), as a function of detected linear polarization. The numbers indicate the value of the fine structure splitting for the X and XX lines. (c) Power-dependence of the μPL intensity for the selected spectral lines. The numbers indicate the exponent in the fitted power function ($I \propto P^\alpha$), which is shown by the solid straight line.

We meticulously tested the possibility of multiphoton generation events for the QD emitter. Thus, single photon emission purity, focusing on the X state, was determined under the non-resonant (660 nm) CW excitation. In Figure 4a), one can see the spectrum of the QD emission after it was coupled to a single-mode fiber, passed through a tunable fiber optic filtering system, and then recorded by an SNSPD. The spectrum was captured at low excitation power conditions ($P_{exc}$<10 μW) and the chosen X emission exhibited a low background counts level on the order of 50 Hz, which is highly desirable for minimizing the contribution of multiphoton coincidences. To measure photon statistics in the HBT interferometer, the X emission was filtered out from the spectrum and transmitted through the 1:2 fiber optic coupler ended by two SNSPDs. The recorded autocorrelation histogram, as shown in Figure 4b), demonstrates high purity of single photon emission, evidenced by the value of the $g(2)(0) = 0.04^{+0.08}_{-0.04}$, evaluated with the fitting procedure at the antibunching deep, while not a single coincidence at zero time delay was recorded within the entire ca. two hours of the histogram collection (see the inset of Fig. 4b).

Next, the purity of a single photon was tested under the pulse excitation regime. The QD was excited by ~50 ps-long optical pulses with a 76 MHz repetition frequency at 805 nm and an average power of 100 μW. The resulting μPL spectrum is presented in Figure 4c), showing ~8 kHz counts of the X emission, ~10 kHz counts of the CX emission and around 1.5 kHz counts for the XX line. Figure 4d) displays the registered autocorrelation histogram in the HBT experiment for the X emission. Here, we observe inevitable background coincidences originating from the background emission seen in the SNSPD spectra on the level of ~800 Hz. Therefore, the evaluation of X emission purity under pulsed excitation takes into account signal (~8kHz) to background (~800Hz) ratio $\rho = s/(s+b) = 0.91$ and the related background $g2(\tau)$ contribution of $1-\rho^2 = 0.17$, which results in background purity of the source of 0.01, while raw data reveals purity of ~0.18.

After estimating the count rates for selected QD emission lines and the background counts under the optical pulse driving, we could estimate extraction efficiency from the FIB-milled mesa. At these excitation conditions (low excitation power), two main competitive recombination channels, X and CX, are observed, with negligible contributions from the XX (see Fig. 4c)). Therefore, using the all-fiber setup, we use the sum of X and CX SNSPD counts to evaluate the photon extraction efficiency ($\eta_{exp}$) from a single QD. This translates to $\eta_{exp} = \frac{18kHz}{76MHz \cdot 0.1\%} = 0.24$. These estimates have been taken with (0.10 ±0.05)% setup efficiency for the NA = 0.65 (see a detailed evaluation above), assuming perfect QD internal extraction efficiency. Apparently, the $\eta_{exp}$ is less than the calculated one (η = 0.40) for this mesa geometry. The difference can be attributed to the QD position shifting from the mesa's center or not having an ideal QD internal quantum efficiency influenced by non-radiative recombination [50,51]. However, the most significant finding is the obtained 24% extraction efficiency from an InAs/InP QD emitting at nearly 1550 nm enclosed within the FIB-fabricated mesa, which presents a record value. This is almost two times overcoming the previously obtained $\eta_{exp}$ of ~13% for a planar QD structure on a distributed Bragg reflector [21] and 17% for a QD coupled to a circular Bragg grating [52][13].

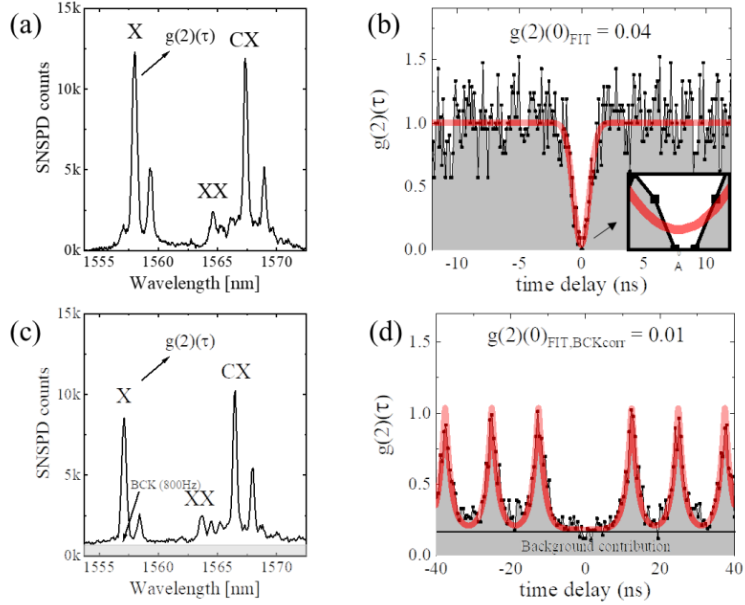

Fig. 4. Quantum dot spectra from a non-deterministic mesa structure realized by Xe-PFIB under (a) continuous wave 660 nm and (b) pulsed 805 nm excitation, and the respective autocorrelation histograms (b,d) showing high suppression of multiphoton events at zero time delay of 0.04 (b) and 0.01 (c)

## 5. Deterministic fabrication of photonic mesa structures

The FIB milling method for making the QD-based devices can be efficiently combined with pre-selection of the QDs from a planar structure by, e.g., a fast optical imaging technique. [12][13]. Here, we show that such a combination paves the way for rapid and cost-effective deterministic fabrication of bright QD-based single photon sources for the example of application-relevant 1550 nm wavelength range. In this case, the planar structure with InAs/InP QDs on silicon was processed in the Xe-PFIB with a high-energy beam to fabricate alignment marks on the structure's surface, defining areas of $20\times30$ $\mu m^2$ grouped into several blocks. The QD preselection requires a few steps, so we collected each field in sequence PL images for 1200 nm long pass (LP) filtering only and for both 1500 nm LP filter and 1550±7.5 nm or 1550±1.5 nm band-pass filter. The respective images are presented in Figs. 5(a-c). First, we clearly identify alignment marks taking advantage of strong emission of 2D layer states [53][37,54] around the 1250 nm wavelength range, which overwhelms the QD emission, resulting in a smooth intensity distribution in the planar region and pronounced light scattering on the markers. Secondly, the spectral filtering above 1500 nm is used to retain QD emission (Fig. 5b and 5c), while the additional bandpass filtering allows to observe only those emitting around the target 1550 nm wavelength. The pronounced 2D Gaussian-like peaks were observed in the given field of view, and its center position with respect to the alignment marks is used for deterministic processing (accuracy of the QD position with respect to alignment marks is ~1 µm, which results from the temporal stability of the setup within the 30 second of measurement). Next, we validate the spectra from the isolated QDs by collecting the µPL on these bright spots using confocal filtering on the 100 µm pinhole. The sample is then processed by Xe-PFIB to fabricate mesas in the specific positions. In this way, we made two mesas on the field with deterministically integrated QDs (QD1 and QD2), as shown in Fig. 5d).

To provide additional insight into the quality of deterministic processing using Xe-PFIB, we plot µPL spectra for each individual mesa before and after fabrication. First, one can notice that in the given spectral range, the emission pattern looks similar for fabricated mesas before and after fabrication, implying no degradation of the crystal structure in the vicinity of the emitter. In the case of QD1, the intensity of specific transitions at 1550 nm is approx. 3 times higher after the processing, which confirms the improved collection efficiency despite a relatively simple photonic microstructure. The spectra are taken for the same excitation power measured in front of the microscope objective, while we observe even more emission lines due to enhanced extraction. Considering the QD2 case, the spectra show no pronounced emission lines in the 1550±1.5 nm range which is consistent with no signature of emission in Fig. 5c). Although some background emission of each QD inherently related to phonon coupling or higher order excitonic states typically gives some features on the PL map (low-intensity spectral profile is integrated, which generates detectable counts with high-sensitivity 2D camera chip on a small region of interest), we can eliminate this effectively adjusting the signal-to-background ratio as a threshold for data visualization to preselect only the dots with pronounced emission lines. We suspect that the positioning accuracy can be improved with a better design of alignment marks, while the control of the FIB processing typically has high accuracy and might not be a critical issue here. Last, the deterministic processing approach is generally required for scalable fabrication purposes, and our results verify its high impact.

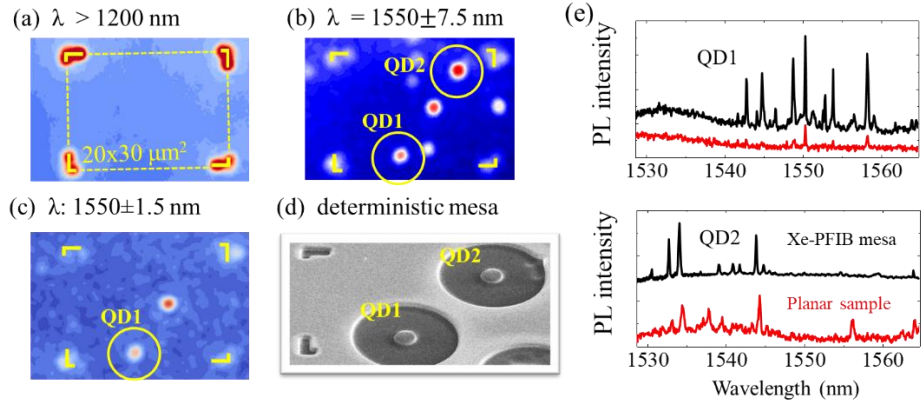

Fig. 5. (a) Photoluminescence wide-field imaging of planar samples using (a) long pass filtering of 1200 nm, (b,c) long pass filtering of 1500 nm together with bandpass filtering centered at 1550 with (b) 15 nm band and (c) 3 nm band. (d) Deterministically fabricated mesas at QD positions found in (b). (e) Microphotoluminescence spectra of QDs were selected from the PL image in (b) before and after mesa processing.

## 6. Conclusions

In this study, we have showcased the unique potential of Xe-PFIB in the manufacturing process of photonic structures, particularly in the prototyping of quantum photonic devices. Our use of this cutting-edge technology has allowed us to fabricate deterministic and non-deterministic mesa structures with a truncated cone shape, featuring enhanced photon extraction efficiency from a single photon emitter operating near 1550 nm and buried in an III-V semiconductor matrix. The photon emitter comprises a self-assembled InAs/InP quantum dot in a hybrid structure with an aluminum mirror at the bottom, all integrated into the silicon platform. For the mesa structure, we have presented FDTD simulations, which indicate a significantly elevated photon extraction efficiency ($\eta$) at the collection lens exceeding 0.6 in the wavelength range between 1.50 to 1.58 µm with the peak efficiency of $\eta = 0.89$ (for NA = 0.65). The experiment provided extraction efficiency of 24% with NA=0.65, which is lower than the calculated one due to imperfections in the system. The QD-based single photon source fabricated through the process exhibited high emission purity, as evidenced by the $g^{(2)}(0)$ function value of 0.04 and 0.01 under continuous wave and pulsed excitation, respectively.

We have shown that Xe-PFIB technology can be effectively combined with fast PL imaging of self-assembled quantum dot structures to create photonic devices with well-controlled parameters. Our photonic devices, which have a pre-selected InAs/InP quantum dot emitting at 1550 nm in a mesa structure, displayed a 3-fold increase in emission intensity compared to a flat, unpatterned system.

**Acknowledgements**

This work has been supported by the "Implementation doctorate" program of the Polish Ministry of Education and Science within grant No. DWD/4/50/2020, the grant no. 2019/33/B/ST5/02941 of OPUS17 call of the National Science Centre in Poland, and grant no. 2020/39/D/ST5/02952 of SONATA16 call of the National Science Centre in Poland.